\documentclass[preprint,3p, twocolumn]{elsarticle}
\usepackage{amsmath,amsfonts,amssymb} 
\usepackage{graphicx,slashed}
\pdfoutput=1

\input xy
\xyoption{all}
\xyoption{curve}

\def\bar{\overline}

\def\cO{\mathcal{O}}

\def\dirac{\slashed{\partial}}

\def\gf{\mathrm{gf}}
\def\gh{\textup{gh}}

\def\k{\mathbf{k}}
\def\L{\mathcal{L}}

\def\R{\mathbb{R}}

\def\su{\mathfrak{su}}

\DeclareMathOperator{\tr}{Tr}

\def\u{\mathfrak{u}}

\begin{document}

\begin{frontmatter}
\title{Renormalizability conditions for almost-commutative geometries}

\author{Walter D. van Suijlekom}
\ead{waltervs@math.ru.nl}

\address{Institute for Mathematics, Astrophysics and Particle Physics,
Radboud University Nijmegen, Heyendaalseweg 135, 6525 AJ Nijmegen, The Netherlands}

\date{18 April 2012}

\begin{abstract}
We formulate conditions for almost-commutative (spacetime) manifolds under which the asymptotically expanded spectral action is renormalizable. These conditions are of a graph-theoretical nature, involving the Krajewski diagrams that classify such geometries.
This applies in particular to the Standard Model of particle physics, giving a graph-theoretical argument for its renormalizability. A promising potential application is in the selection of physical (renormalizable) field theories described by almost-commutative geometries, thereby going beyond the Standard Model. 

\end{abstract}

\end{frontmatter}


Over the past few years it has turned out that many particle physics models can be described geometrically by modifying the internal structure of spacetime, making it slightly noncommutative. 
Indeed, these so-called {\it almost-commutative geometries} allow for a (geometrical) 
derivation of Yang--Mills theory \cite{CC96}, or even the full Standard Model, including Higgs potential and neutrino mass terms \cite{CCM07, CM07, JKSS07}, all minimally coupled to gravity. Theories that go beyond the Standard Model were described in \cite{Ste06,Ste07,SS07,JS08}, while some supersymmetric models 
have been derived geometrically in \cite{BroS11}. 
The basic idea in all these examples is that one describes an almost-commutative geometry by spectral data, and then applies a general spectral action principle to derive physical Lagrangians. 

In this Letter we 
try to understand 
the perturbative renormalizability of these Lagrangians. Focusing on the gauge theoretical part, we avoid the non-renormalizable gravitational background, thus parting from the unified picture provided by noncommutative geometry. Nevertheless, we note the intriguing appearance of higher-derivative terms in the gravitational sector as well \cite{CCM07}, potentially resolving this non-renormalizability problem \cite{Ste77}. 

We will formulate graph-theoretical conditions for almost-commutative geometries that render the spectral action (at lowest order in a cutoff) renormalizable as a gauge theory. 
This generalizes our previous result on (super)renormalizability of the asymptotically expanded Yang--Mills spectral action \cite{Sui11b,Sui11c} to a more general class of particle physics models. 
In particular, it gives a graph-theoretical proof of renormalizability  of the Standard Model. 

For a more detailed treatment of these results, we refer to the companion preprint \cite{Sui11d}. 

\section{Matrices and particle physics}
Let us introduce the noncommutative spacetimes that are of interest in describing particle physics models \cite{CC96,CCM07,CM07}, which are referred to in the literature as {\it almost-commutative} geometries. Indeed, the noncommutativity we will encounter here is rather mild, and is related to the ordinary matrix product. We stress that this noncommutativity is not of the Moyal-type, where canonical commutation relations are introduced between the spacetime coordinates. 

Essentially, we describe Kaluza--Klein-like spacetimes
$
M \times F,
$
which are a product of spacetime $M$ with a finite {\it noncommutative space} $F$. The space $F$ is noncommutative in the sense that its coordinates are {\it matrix-valued}. In other words, at each point $x$ of spacetime $M$ we consider $N$-tuples $a(x) = (a_1 (x), \ldots, a_N(x))$ of (real, complex, or quaternion-valued) square matrices $a_i$, say of size $k_i \times k_i$. 


\subsection{Fermionic fields}
Given the above matrices $a_i$ at each point of spacetime $M$, we now define fermionic fields. Adopting the fundamental idea that particles are representations, we define a fermionic field as a vector on which the above matrices act by matrix multiplication. This can be from the left or from the right, considering the vector as a column vector or as a row vector, respectively. We choose to consider both at the same time, so that a basic fermionic constituent $\psi$ of our theory is a tensor product 
$$
\psi = \begin{pmatrix} v_1 \\ \vdots \\ v_{k_i} \end{pmatrix} \otimes \begin{pmatrix} w_1 & \cdots &w_{k_{i'}} \end{pmatrix}.
$$
Then, the $k_i \times k_i$ matrix $a_i$ acts on the first vector from the left, and the $k_{i'} \times k_{i'}$ matrix $a_{i'}$ acts on the second vector from the right. As usual, we indicate this representation by their dimensions, written in bold as ${\bf k_i} \otimes {\bf k_{i'}}$. 

A convenient diagrammatic way to express this is as follows. First, we label horizontal and vertical axes by the given integers ${\bf k_1}, \ldots , {\bf k_N}$. Then, we indicate the presence of a fermion $\psi$ in the left representation of $a_i$ and in the right representation of $a_{i'}$ by a vertex at position $({\bf k_i},{\bf k_{i'}})$ (cf. Figure \ref{fig:kra}).



\subsection{Bosonic fields}
In the noncommutative description of particle physics models all bosonic fields (scalar and gauge) nicely arise in the same way. Essentially, the scalar fields can be interpreted as gauge fields in the finite noncommutative space $F$. Let us make this more precise, using the diagrammatic approach of the previous subsection. 

\begin{figure}[t]
\begin{center}
\begin{tabular}{c}
\begin{xy} 0;<3mm,0mm>:<0mm,3mm>::0;0,
,(5,0)*{\cdots}
,(8,0)*{\k_i}
,(11,0)*{\cdots}
,(14,0)*{\k_{j}}
,(17,0)*{\cdots}
,(1.5,-2.5)*{\vdots}
,(1.5,-5.5)*{\k_{i'}}
,(1.5,-8.5)*{\vdots}
,(1.5,-11.5)*{\k_{j'}}
,(1.5,-14.5)*{\vdots}
,(8,-5.5)*\cir(0.3,0){}
,(14,-5.5)*\cir(0.3,0){}
,(8,-11.5)*\cir(0.3,0){}
,(8,-5.5)*\cir(0.4,0){}
,(8.3,-5.5);(13.8,-5.5)**\dir{-}
,(8,-5.8);(8,-11.3)**\dir{-}
\end{xy}
\end{tabular}
\caption{The vertices (possibly doubled indicating multiplicities) represent fermionic fields; an edge between two vertices represents a scalar field $\phi$ and its conjugate $\phi^\dagger$. }
\label{fig:kra}
\end{center}
\end{figure}
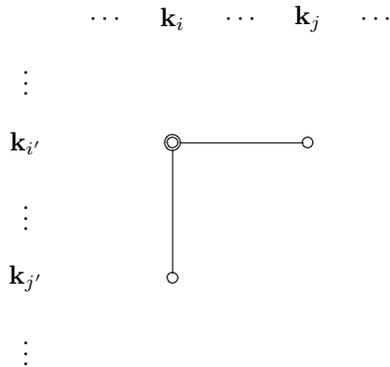

We define a bosonic field simply as a certain linear map between fermions. Thus, in terms of our diagrams, bosonic fields map between vertices in the diagram. 
If the initial and final vertex are different, say, ${\bf k_i} \otimes {\bf k_{i'}}$ and ${\bf k_{j}} \otimes {\bf k_{i'}}$, the bosonic field is a ${\bf k_i} \times{\bf k_{j}}$ matrix. Its hermitian conjugate $\phi^\dagger$ then gives a map in the opposite direction. We indicate both maps $\phi$ and $\phi^\dagger$ in the diagram by a single edge, as in Figure \ref{fig:kra}. 

The matrix $\phi$ will be called a {\it scalar field}, which is for good physical reasons as we will see below. The fields $\phi$ at the edges are collected into one scalar field $\Phi$. By construction, it is hermitian $\Phi^\dagger = \Phi$.

If the initial and final vertex of such a linear map coincide, we can also differentiate the fermionic field with respect to the spacetime coordinates $x^\mu$. In this case, we obtain the covariant derivative combined with Clifford multiplication, acting on the fermion at that vertex as
$i \gamma^\mu \circ \nabla_\mu$ with $\nabla_\mu \psi = \partial_\mu \psi - i (A_\mu \cdot \psi - \psi \cdot A_\mu)$ and with $\gamma^\mu$ the Dirac gamma matrices. We will call $A_\mu$ the {\it gauge field}. For each $\mu$ it is a matrix of the same form as the $a(x)$ above. We require the gauge field $A_\mu$ to be real 
$A_\mu(x) ^\dagger = A_\mu(x)$, and moreover traceless:
$ \tr A_\mu(x)  =0 .
$
The trace is taken over all fermions; this is the unimodularity condition on gauge fields.


The above diagrams were first introduced by Krajewski \cite{Kra97} and are therefore called {\it Krajewski diagrams}; they classify all possible noncommutative spaces $F$. In addition to the above rules of drawing only horizontal and vertical lines, they should be symmetric along the diagonal, corresponding to the symmetry between particles and anti-particles. Around the same time, such a classification was obtained also in \cite{PS98}.

\subsection{Gauge transformations}
Let us continue the above line of {\it deriving} particle physics models from the matrices we started with. In fact, we can restrict to skew-hermitian matrices $a(x)^\dagger = - a(x)$ and impose that $\tr a(x)= 0 $, to obtain the (generators of) {\it gauge transformations}. Again, the trace here is taken over all fermions, on which $a$ acts on by left matrix multiplication. 
As usual, we will refer to this trace-free condition as the {\it unimodularity} condition, also discussed in the context of anomaly cancellation in \cite{AGM95}.

Since a gauge transformation is still a matrix, it acts on the fermions as
$\psi \mapsto a \cdot \psi - \psi \cdot a,$
combining left and right matrix multiplication. 
It also acts on the bosons by conjugation; 
on the gauge fields we have $\nabla_\mu  \mapsto \nabla_\mu a - a \nabla_\mu$, which is the same as the familiar gauge transformation 
$A_\mu \mapsto \partial_\mu a + A_\mu \cdot a - a \cdot A_\mu$.

\begin{figure}
\begin{center}
\begin{tabular}{c}
\begin{xy} 0;<3mm,0mm>:<0mm,3mm>::0;0,
,(5,-0)*{\cdots}
,(8,-0)*{\k_i}
,(11,-0)*{\cdots}
,(14,-0)*{\k_j}
,(17,-0)*{\cdots}
,(8,-2)*\cir(0.3,0){}
,(14,-2)*\cir(0.3,0){}
,(8.2,-2);(13.8,-2)**\dir{-}
\end{xy}
\end{tabular}
\caption{The horizontal projection of the Krajewski diagram of Figure \ref{fig:kra}.}
\label{fig:kra-proj}
\end{center}
\end{figure}
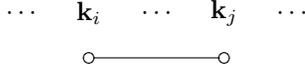

For the scalar fields, it is useful to use a horizontal (or vertical) projection of the Krajewski diagram. Indeed, a scalar field corresponding to, say, the horizontal line in Figure \ref{fig:kra} is a ${\bf k_i \times k_j}$ matrix. As such, it can also be indicated as a line in the projected diagram (Figure \ref{fig:kra-proj}).

Since each vertex in the projected diagram corresponds to a (column) vector, a gauge transformation $a = - a^\dagger$ acts on these vectors. Consequently, there is an action of $a$ on the scalar fields by conjugation: 
$\phi \mapsto a \cdot \phi - \phi \cdot a$.
Indeed, both $a$ and $\phi$ are matrices that can be multiplied.

\subsubsection{Example: The Standard Model}
The Standard Model of high-energy physics can be derived from an almost-commutative geometry \cite{CCM07, CM07, JKSS07}, as we will now briefly recall. 
The matrix $a$ has three components,
\begin{itemize}
\item $z$, a complex $1 \times 1$ matrix, {\it i.e.} a complex number;
\item $q$, a quaternion, written as a $2 \times 2$ matrix $q= q^0 + \sum_\alpha q^\alpha \sigma_\alpha$ in terms of Pauli matrices;
\item $m$, a complex $3 \times 3$ matrix.
\end{itemize}
The fermionic content is described by indicating vertices in a Krajewski diagram (Figure \ref{fig:kra-sm}). The bar in ${\bf \bar 1}$ indicates that the complex number $z$ acts on this fermion by multiplication with its conjugate $\bar z$. Moreover, the ${\bf 2}$ indicates that the quaternion acts on it as a $2 \times 2$ matrix ({\it i.e.} by left multiplication).

\begin{figure}[b]
\begin{center}
\begin{tabular}{c}
\begin{xy} 0;<3mm,0mm>:<0mm,3mm>::0;0,
,(3,0)*{{\bf 1}}
,(6,0)*{{\bf \bar 1}}
,(9,0)*{{\bf 2}}
,(12,0)*{{\bf 3}}
,(1.5,-3)*{{\bf 1}}
,(1.5,-6)*{{\bf \bar 1}}
,(1.5,-9)*{{\bf 2}}
,(1.5,-12)*{{\bf 3}}
,(3,-3)*\cir(0.3,0){}
,(6,-3)*\cir(0.3,0){}
,(9,-3)*\cir(0.3,0){}
,(6.2,-3);(8.8,-3)**\dir{-}
,(3.3,-2.9);(8.8,-2.9)**\crv{(6,-2)}
,(3,-12)*\cir(0.3,0){}
,(6,-12)*\cir(0.3,0){}
,(9,-12)*\cir(0.3,0){}
,(6.2,-12);(8.8,-12)**\dir{-}
,(3.2,-12.1);(8.8,-12.1)**\crv{(6,-13)}
,(3,-3)*\cir(0.4,0){}
,(3,-6)*\cir(0.3,0){}
,(3,-9)*\cir(0.3,0){}
,(3,-6.2);(3,-8.8)**\dir{-}
,(2.9,-3.3);(2.9,-8.8)**\crv{(2,-6)}
,(12,-3)*\cir(0.3,0){}
,(12,-6)*\cir(0.3,0){}
,(12,-9)*\cir(0.3,0){}
,(12,-6.2);(12,-8.8)**\dir{-}
,(12.1,-3.2);(12.1,-8.8)**\crv{(13,-6)}
\end{xy}
\end{tabular}
\caption{The Krajewski diagram of the Standard Model}
\label{fig:kra-sm}
\end{center}
\end{figure}
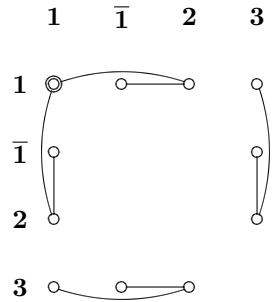

\begin{figure}
\begin{center}
\begin{tabular}{c}
\begin{xy} 0;<3mm,0mm>:<0mm,3mm>::0;0,
,(6,0)*{{\bf 1}}
,(9,0)*{{\bf \bar 1}}
,(12,0)*{{\bf 2}}
,(15,0)*{{\bf 3}}
,(6,-2)*\cir(0.3,0){}
,(9,-2)*\cir(0.3,0){}
,(12,-2)*\cir(0.3,0){}
,(15,-2)*\cir(0.3,0){}
,(9.2,-2);(11.8,-2)**\dir{-}
,(6.2,-1.9);(11.8,-1.9)**\crv{(9,-1)}
\end{xy}
\end{tabular}
\caption{The projection of the Krajewski diagram of the Standard Model of Figure \ref{fig:kra-sm}.}
\label{fig:kra-sm-proj}
\end{center}
\end{figure}
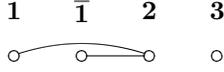

Let us work through the particle content of this Krajewski diagram. 
Consider one of the two vertices on the top left corner in the diagram.  It describes a fermion, which we will denote by $\nu_R$, in the ${\bf 1 \otimes 1}$. Recall that a gauge transformation is a skew-hermitian matrix $a =(z, q, m)$, so that we have $\bar z = -z, q^\dagger = - q$ and $m^\dagger = - m$. 
This means that $z$ is of the form $i t$, with $t \in \R$; {\it i.e.} $z$ is an element in the Lie algebra $\u(1)$ which we will soon relate to hypercharge. A gauge transformation acts on $\nu_R$ as 
\begin{align*}
\nu_R &\mapsto a \cdot \nu_R - \nu_R \cdot a= it \nu_R - it \nu_R  = 0.
\intertext{Let us then consider the next vertex on this row, which is in the ${\bf \bar 1 \otimes 1}$ and will be denoted by $e_R$. A gauge transformation now acts as:}
e_R &\mapsto a \cdot e_R - e_R \cdot a = -i t e_R - i t e_R = - 2it e_R.
\end{align*}
We interpret the $0$ and $-2$ as the hypercharge of $\nu_R$ and $e_R$, respectively. 

Next, more interestingly, we find a vector ${\bf 2 \otimes 1}$ on the top row, which we will denote by ${\bf L}$. If we note that a skew-hermitian quaternion $q$ is nothing but an $\su(2)$-matrix acting on the ${\bf 2}$, a gauge transformation acts on ${\bf L}$ as 
$$
{\bf L} \mapsto a \cdot {\bf L} - {\bf L} \cdot a = \sum_\alpha q^\alpha \sigma_\alpha \cdot {\bf L} - i t {\bf L}.
$$
Indeed, the ${\bf 2 \otimes 1}$ denotes a left representation of $q$, and a right representation of $i t$. 
In other words, ${\bf L}$ is an $\su(2)$-doublet with (hyper)charge $-1$. We conclude that this vertex represents the left electron neutrino and electron.


Before explaining the remaining vertices of the diagram, we note that the unimodularity condition $\tr a = 0$ implies that $a$ is of the form 
$
a = (z, q, m - \frac13 z 1_3)
$
where $1_3$ is the $ 3 \times 3$ identity matrix and now $m \in \su(3)$. 
Let us consider how this element acts on the vertices on the bottom row. From left to right, we first encounter a ${\bf 1 \otimes 3}$, followed by a ${\bf \bar 1 \otimes 3}$. We denote the two fermionic fields by $u_R$ and $d_R$, respectively. A gauge transformation then acts as
\begin{align*}
u_R & \mapsto a \cdot u_R - u_R \cdot a = \tfrac{4}{3} it u_R - u_R \cdot m, \\
d_R & \mapsto a \cdot d_R - d_R \cdot a =- \tfrac{2}{3} it d_R - d_R \cdot m.
\end{align*}
Thus, $u_R$ and $d_R$ are the right up and down quark with hypercharges $+4/3$ and $-2/3$, respectively. Note that these hypercharges are not imposed but follow from the unimodularity condition on $a$.

The remaining vertex on the bottom row is ${\bf 2} \otimes {\bf 3}$. We denote the corresponding fermionic field wishfully by ${\bf Q}$, and find that a gauge transformation acts as
$$
{\bf Q} \mapsto a \cdot {\bf Q} - {\bf Q} \cdot a= q \cdot {\bf Q} - {\bf Q} \cdot m +\frac13 z {\bf Q}
$$
From this, we read off that ${\bf Q}$ is an $\su(2)$-doublet, an $\su(3)$-triplet, and has hypercharge $1/3$, thus describing the left up and down quark. 

The remaining vertices in the diagram correspond to the respective anti-particles. Three generations can be taken into account by tripling all the vertices in the Krajewski diagram. 

A completely analogous analysis leads to the correct coupling of the gauge fields to fermions. Since gauge fields $A_\mu$ are given by hermitian, traceless elements in the matrix algebra, they take the form
$
A_\mu = (B_\mu, W_\mu, V_\mu - \frac13 B_\mu 1_3)
$
This traceless form automatically implies that $B_\mu$ acts on the fermions according to the right hypercharges. Moreover, the gauge fields transform in the adjoint representation of $\u(1) \oplus \su(2) \oplus \su(3)$, as desired. 

As far as the scalar fields are concerned, we project the diagram to obtain the horizontal projection in Figure \ref{fig:kra-sm-proj}. The field $\phi$ appearing in this diagram is a $2 \times 1$ matrix $\phi = \begin{pmatrix} \phi_1 \\ \phi_2 \end{pmatrix}$; it is the 
Higgs doublet with hypercharge $-1$. It is coupled to all the different fermion flavours and generations according to the Yukawa mass matrix. An interesting noncommutative geometrical perspective on the mixing of flavour through the CKM-matrix is given in \cite{C08b}.

\subsection{Spectral action}
We return again to the general construction, whilst keeping the Standard Model as a guiding example in what follows.

In order to describe the dynamics and interactions for the above field content, we need a Lagrangian for the fermionic and bosonic fields $\psi$, $A_\mu$ and $\phi$. Naturally, we want this Lagrangian to be invariant under the gauge transformations described 
above. We will adopt a {\it spectral} point of view and count eigenvalues of a generalized Dirac operator on noncommutative spacetime $M \times F$. 

This Dirac operator is defined by combining the usual Dirac operator on $M$ with a `finite' Dirac operator on $F$, related to the scalar fields $\phi$. In other words, we set
$
\dirac_{M \times F}  = i \gamma^\mu \nabla_\mu + \gamma^5 \Phi.
$ 
This generalizes the minimally coupled (to $A_\mu$) Dirac operator on $M$ to an operator on $M \times F$. The action of $\dirac_{M \times F}$ is according to the representation of the fields $A_\mu$ and $\phi$ on the fermions, as described above.


The {\it spectral action} is given by the number of eigenvalues of $\dirac_{M \times F}$ that are smaller (in absolute value) than a given cutoff $\Lambda$. If we consider this as a functional of the field $A_\mu$ and $\phi$ and assume a smooth cutoff
, we obtain after some computations \cite{CCM07,CM07} the following Lagrangian 
$$
\L_{M \times F} = \sum_{n\geq 0} \Lambda^{4-m} f_{4-m}  a_m(x, \dirac_{M\times F}^E).
$$
as an asymptotic expansion for large $\Lambda$. Here $f_{4-m}$ are moments of the cutoff function that was used, and $a_m$ are the so-called heat invariants -- a technique exploited already by Schwinger in \cite{Sch51} -- of the generalized Dirac operator (in Euclidean signature). Explicit expressions for the first few heat invariants can be found in eg. \cite{Avr91,Vas03}, leading on a flat background $M$ to
\begin{multline}
\label{eq:sa}
\L_{M \times F} = \frac{f_4 N \Lambda^4}{2 \pi^2} - \frac{f_2\Lambda^2}{2 \pi^2} \tr \Phi^2\\
+ \frac{f_0}{8 \pi^2}  \tr \left( (\nabla_\mu \Phi)^2 + \Phi^4 \right) \\
  - \frac{f_0}{24 \pi^2} \tr F_{\mu \nu} F^{\mu\nu}  + \cO(\Lambda^{-1})
\end{multline}
This is our (Euclidean) Lagrangian for the scalar and gauge fields, written in terms of the field strength $F_{\mu\nu}$ of the gauge fields. Being spectrally defined, it is manifestly gauge invariant.
We consider the terms that are proportional to inverse powers of $\Lambda$ as being suppressed by this large cutoff and focus on the first few terms. 

The terms proportional to $\tr \Phi^2$ and $\tr \Phi^4$ can be nicely described in terms of loops in the Krajewski diagram. In fact, the non-zero contributions to $\tr \Phi^2$ come from going back-and-forth an edge in the Krajewski diagram, corresponding to the action of a scalar field component $\phi$ composed with its adjoint $\phi^\dagger$. The corresponding term in the Lagrangian is $\tr \phi^\dagger \phi$. 

\begin{figure}
\begin{center}
\vspace{-15mm}
\begin{tabular}{c}
\begin{xy} 0;<3mm,0mm>:<0mm,3mm>::0;0,
,(-5,-7.5)*{{\bf (I)}}
,(-5,-13.5)*{{\bf (II)}}
,(0,-7.5)*\cir(0.3,0){}
,(2,-7.5)*\cir(0.3,0){}
,(4,-7.5)*\cir(0.3,0){}
,(6,-7.5)*\cir(0.3,0){}
,(0.2,-7.5);(1.8,-7.5)**\dir{-} ?>*\dir{>}
,(2.2,-7.5);(3.8,-7.5)**\dir{-} ?>*\dir{>}
,(4.2,-7.5);(5.8,-7.5)**\dir{-} ?>*\dir{>}
,(0.2,-7.3);(5.8,-7.3)**\crv{(3,-5)} ?<*\dir{<}
,(10,-7.5)*\cir(0.3,0){}
,(13,-7.5)*\cir(0.3,0){}
,(16,-7.5)*\cir(0.3,0){}
,(10.2,-7.5);(12.8,-7.5)**\crv{(11.5,-8.5)} ?>*\dir{>}
,(13.2,-7.5);(15.8,-7.5)**\crv{(14.5,-8.5)} ?>*\dir{>}
,(10.2,-7.5);(12.8,-7.5)**\crv{(11.5,-6.5)} ?<*\dir{<}
,(13.2,-7.5);(15.8,-7.5)**\crv{(14.5,-6.5)} ?<*\dir{<}
,(2,-12)*\cir(0.3,0){}
,(5,-12)*\cir(0.3,0){}
,(2.2,-12);(4.8,-12)**\crv{(3.5,-11)} ?<*\dir{<}
,(2.2,-12);(4.8,-12)**\crv{(3.5,-13)} ?>*\dir{>}
,(2,-15)*\cir(0.3,0){}
,(1.8,-12);(1.8,-15)**\crv{(1,-13.5)} ?<*\dir{<}
,(2.2,-12);(2.2,-15)**\crv{(3,-13.5)} ?>*\dir{>}
,(12,-12)*\cir(0.3,0){}
,(15,-12)*\cir(0.3,0){}
,(12,-15)*\cir(0.3,0){}
,(15,-15)*\cir(0.3,0){}
,(12.2,-12);(14.8,-12)**\dir{-} ?>*\dir{>}
,(15,-12.2);(15,-14.8)**\dir{-} ?>*\dir{>}
,(12.2,-15);(14.8,-15)**\dir{-} ?<*\dir{<}
,(12,-12.2);(12,-14.8)**\dir{-} ?<*\dir{<}
\end{xy}
\end{tabular}
\end{center}
\caption{Possible loops that can appear in a Krajewski diagram of length 4, giving rise to quartic scalar couplings. We distinguish two types of loops: loops along a single row or column ({\bf I}), and mixed (horizontal and vertical) loops ({\bf II}).}
\label{fig:loops4}
\end{figure}
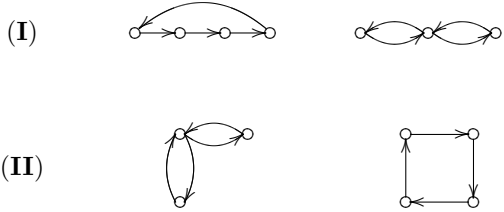

Similarly, the term $\tr \Phi^4$ corresponds to loops in the Krajewski diagram of length $4$. Such loops arise in different forms, as depicted in Figure \ref{fig:loops4}. For each such loop, the resulting term in the Lagrangian is quartic in the scalar fields associated to the four edges of that loop. 
\begin{figure}
\begin{center}
\vspace{-15mm}
\begin{tabular}{c}
\begin{xy} 0;<3mm,0mm>:<0mm,3mm>::0;0,
,(20,-7.5)*\cir(0.3,0){}
,(25,-7.5)*\cir(0.3,0){}
,(20.2,-7.5);(24.8,-7.5)**\crv{(22.5,-6)} ?>*\dir{>}
,(20.2,-7.5);(24.8,-7.5)**\crv{(22.5,-9)} ?<*\dir{<}
\end{xy}
\end{tabular}
\end{center}
\caption{Loops in a Krajewski diagram of length 2 are necessarily given by going back-and-forth a single edge.}
\label{fig:loops2}
\end{figure}
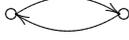

For the fermions, the most natural candidate Lagrangian is
\begin{equation}
\label{eq:ferm}
\L_f = \bar \Psi \dirac_{M \times F} \Psi \equiv i  \bar \Psi \gamma^\mu \nabla_\mu \Psi + \bar \Psi \gamma^5 \Phi \Psi,
\end{equation}
where we have collected all fermionic fields $\psi$ (at each vertex of the diagram) into a single fermion $\Psi$. A consequence of working with the Euclidean signature is that $\bar \Psi$ should be defined as $(J \Psi)^\dagger$. This is explained in full detail in \cite{CCM07}, and is confronted with the Lorentizan model in \cite{Bar06}.

For the noncommutative description of the Standard Model the Lagrangians $\L_{M \times F}$ and $\L_f$ indeed reproduce the full Lagrangian of the Standard Model, including Higgs mechanism \cite{CCM07}.

\section{Renormalization for almost-commutative geometries}
Our main result is the formulation of conditions on $F$ to make the above Lagrangian renormalizable. These conditions are of graph-theoretical nature: they can nicely be expressed in terms of the Krajewski diagram underlying the field theory. 

First, we add a gauge fixing term to the above Lagrangian. Given the presence of a Higgs-type potential for the scalar fields $\Phi$, we introduce a 't Hooft's $R_\xi$-gauge: 
$$
\L_\gf [A,\Phi] = f_0 \tr \left (\partial_\mu A^{a\mu} - \xi \chi [T^a , v] \right)^2
+ \cO(\Lambda^{-1})
$$
in terms of $A^\mu = A^{a\mu} T^a$ and expanding $\Phi = v + \chi$ around its vacuum-expectation value $v$. The corresponding Faddeev--Popov ghost Lagrangian is then given by
\begin{multline*}
\L_\gh [A,\bar C, C,\Phi] = -f_0 \tr \big( \bar C^a \partial_\mu \partial^\mu C^a \\
+  \bar C^a \partial_\mu [A^\mu,C]^a + \xi \bar C^a [C,\Phi][T^a,v] \big) + \cO(\Lambda^{-1}).
\end{multline*}
The trace is still over all fermions. For more details, and also explicit expressions for the terms proportional to $\Lambda^{-1}$, we refer to \cite{Sui11d}.

After the above gauge-fixing the Lagrangian $\L_{M \times F}$ is power-counting renormalizable if we neglect terms proportional to $\Lambda^{-1}$. 
Since we are dealing with a gauge theory, it is crucial to maintain gauge invariance in the process of renormalization. As a matter of fact, we have to guarantee that the counterterms are of the same form as the monomials appearing in Eq. \eqref{eq:sa} and \eqref{eq:ferm}. Using gauge-invariance of the counterterm, we now determine their general form:

\medskip 

\noindent {\bf gauge fields:} the only gauge invariant expressions in the gauge fields is the Yang--Mills Lagrangian,
$
\tr F_{\mu\nu}F^{\mu\nu}.
$
Note that different factors of the gauge algebra might give rise to different pre-factors for the corresponding Yang--Mills term. In any case, all these terms are already present in Eq. \eqref{eq:sa} and can thus be renormalized. 

\medskip

\noindent{\bf scalar kinetic terms: } the counterterms are the minimally coupled
$
\tr \nabla_\mu \phi^\dagger \nabla^\mu \phi
$
for each scalar field $\phi$. These terms are already present in Eq. \eqref{eq:sa}, albeit that they appear there with the same pre-factor for each field $\phi$.

\medskip

\noindent{\bf quadratic scalar terms:} The counterterms that are quadratic in $\phi$ are 
$
\tr \phi^\dagger \phi
$, already present in Eq. \eqref{eq:sa}.

\medskip

\noindent {\bf quartic scalar terms:} In order to determine all the gauge invariant counterterms that are of order 4 in the scalar fields, the projected diagram (cf. Figure \ref{fig:kra-proj}) is quite useful. We already mentioned that each scalar field $\phi$ corresponds to an edge in this horizontal projection, and that a gauge transformation acts by conjugation on the matrix $\phi$. Then, any gauge-invariant expression in the $\phi$'s (no derivatives) is given by taking the trace of a product of matrices $\phi$ that correspond to a loop in the projected diagram. We have seen one of them already in the previous case: a loop of length two going back-and-forth a single edge. At order four, there are more possibilities, as depicted in Figure \ref{fig:loops4-proj}. For $\L_{M \times F}$ to be renormalizable, the monomials corresponding to these loops must be already present in $\L_{M \times F}$, leading to the following renormalizability condition:

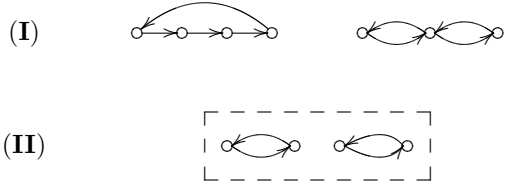
\begin{figure}
\begin{center}
\vspace{-15mm}
\begin{tabular}{c}
\begin{xy} 0;<3mm,0mm>:<0mm,3mm>::0;0,
,(-5,-7.5)*{{\bf (I)}}
,(-5,-12.5)*{{\bf (II)}}
,(0,-7.5)*\cir(0.3,0){}
,(2,-7.5)*\cir(0.3,0){}
,(4,-7.5)*\cir(0.3,0){}
,(6,-7.5)*\cir(0.3,0){}
,(0.2,-7.5);(1.8,-7.5)**\dir{-} ?>*\dir{>}
,(2.2,-7.5);(3.8,-7.5)**\dir{-} ?>*\dir{>}
,(4.2,-7.5);(5.8,-7.5)**\dir{-} ?>*\dir{>}
,(0.2,-7.3);(5.8,-7.3)**\crv{(3,-5)} ?<*\dir{<}
,(10,-7.5)*\cir(0.3,0){}
,(13,-7.5)*\cir(0.3,0){}
,(16,-7.5)*\cir(0.3,0){}
,(10.2,-7.5);(12.8,-7.5)**\crv{(11.5,-8.5)} ?>*\dir{>}
,(13.2,-7.5);(15.8,-7.5)**\crv{(14.5,-8.5)} ?>*\dir{>}
,(10.2,-7.5);(12.8,-7.5)**\crv{(11.5,-6.5)} ?<*\dir{<}
,(13.2,-7.5);(15.8,-7.5)**\crv{(14.5,-6.5)} ?<*\dir{<}
,(4,-12.5)*\cir(0.3,0){}
,(7,-12.5)*\cir(0.3,0){}
,(9,-12.5)*\cir(0.3,0){}
,(12,-12.5)*\cir(0.3,0){}
,(4.2,-12.5);(6.8,-12.5)**\crv{(5.5,-13.5)} ?>*\dir{>}
,(9.2,-12.5);(11.8,-12.5)**\crv{(11,-13.5)} ?>*\dir{>}
,(4.2,-12.5);(6.8,-12.5)**\crv{(5.5,-11.5)} ?<*\dir{<}
,(9.2,-12.5);(11.8,-12.5)**\crv{(11,-11.5)} ?<*\dir{<}
,(3,-11);(13,-11)**\dir{--}
,(3,-11);(3,-14)**\dir{--}
,(3,-14);(13,-14)**\dir{--}
,(13,-11);(13,-14)**\dir{--}
\end{xy}
\end{tabular}
\end{center}
\caption{Loops of length 4 in a projected Krajewski diagram give rise to quartic gauge-invariant counterterms given by traces along these loops, again appearing in two types: single loops ({\bf I}), and pairs of loops ({\bf II}). 
For the loops of type {\bf II}, we exclude the case that they are connected to a common vertex ${\bf 1}$ or ${\bf \bar 1}$. 
}
\label{fig:loops4-proj}
\end{figure}

\medskip

\noindent {\it {\bf (R1)} any loop of length four in the projected Krajewski diagram (Fig. \ref{fig:loops4-proj}) can be lifted to a loop of length four of the same type in the Krajewski diagram (Fig. \ref{fig:loops4}).}

\medskip

\medskip

\noindent{\bf fermionic kinetic terms:} the possible gauge-invariant counterterms are proportional to
$
i \bar \psi \gamma^\mu \nabla_\mu \psi
$
for each component $\psi$. It is already present in Eq. \eqref{eq:ferm}.

\medskip

\noindent {\bf scalar--fermion interaction terms:} the possible counterterms are of the form
$
\bar \psi \gamma_5 \phi \psi'
$
where the field $\phi$ corresponds to an edge between two vertices in the projected Krajewski diagram. Gauge-invariance demands that the fields $\psi$ and $\psi'$ project to precisely these two vertices. 
This gives us the second renormalization condition:

\medskip

\noindent {\it {\bf (R2)} if any two vertices in the Krajewski diagram project onto two vertices of the projected Krajewski diagram which are connected by an edge, then they are connected by an edge themselves. }

\medskip

We can now formulate our main result: the field theory defined by $\L_{M \times F} + \L_f$ (cf. Eq. \eqref{eq:sa} and \eqref{eq:ferm}) is renormalizable if the conditions {\bf R1} and {\bf R2} are satisfied, and if no gauge anomalies occur. 
The cancellation of such anomalies has been discussed in terms of Krajewski diagrams already in \cite{Kra97}, imposing further constraints on the diagrams.

As an example, consider once again the noncommutative description of the Standard Model, with its Krajewski diagram given in Figure \ref{fig:kra-sm}. Any loop of type {\bf I} in the projected diagram (Figure \ref{fig:kra-sm-proj}) can indeed be lifted to a horizontal or vertical loop in the Krajewski diagram. There are no loops of type {\bf II} since these are all connected to a common vertex ${\bf 1}$ or ${\bf \bar 1}$. This graphically establishes that the Standard Model is renormalizable. 
Note that condition ${\bf R2}$ guarantees that if there is a Higgs field coupling to the leptons, then it also couples to the quark fields. The cancellation of anomalies for the noncommutative description of the Standard Model is guaranteed precisely by the unimodularity condition on the gauge field, as was shown in \cite{AGM95}.

As an example of a model that does not satisfy the above renormalizability conditions, consider the fermonic field content described by Figure \ref{fig:nonren}. It is somewhat based on the Standard Model, but
the scalar fields are a $1 \times 1$ and a $2 \times 3$ matrix, let us denote them by $\phi$ and $\phi' $, respectively. Then, a possible gauge invariant expression that can appear as a counterterm but never appears as $\tr \Phi^4$ in Eq. \eqref{eq:sa} is
$$
\phi^\dagger \phi
\tr (\phi')^\dagger \phi'.
$$ 
We see that this model does not satisfy condition {\bf R1}: in the projected diagram (Fig. \ref{fig:nonren-proj}) there is a pair of loops of type {\bf II} which does not lift to a single loop of type {\bf II} in the Krajewski diagram (Fig. \ref{fig:nonren}).

\begin{figure}[t]
\begin{center}
\begin{tabular}{c}
\begin{xy} 0;<3mm,0mm>:<0mm,3mm>::0;0,
,(3,0)*{{\bf 1}}
,(6,0)*{{\bf \bar 1}}
,(9,0)*{{\bf 2}}
,(12,0)*{{\bf 3}}
,(1.5,-3)*{{\bf 1}}
,(1.5,-6)*{{\bf \bar 1}}
,(1.5,-9)*{{\bf 2}}
,(1.5,-12)*{{\bf 3}}
,(3,-3)*\cir(0.3,0){}
,(6,-3)*\cir(0.3,0){}
,(9,-3)*\cir(0.3,0){}
,(9.2,-3);(11.8,-3)**\dir{-}
,(3.3,-3);(5.8,-3)**\dir{-}
,(3,-12)*\cir(0.3,0){}
,(3,-3)*\cir(0.4,0){}
,(3,-6)*\cir(0.3,0){}
,(3,-9)*\cir(0.3,0){}
,(3,-3.3);(3,-5.8)**\dir{-}
,(3,-9.2);(3,-11.8)**\dir{-}
,(12,-3)*\cir(0.3,0){}
\end{xy}
\end{tabular}
\caption{An example of a noncommutative space $F$ that does not satisfy the renormalizability condition {\bf R1}.}
\label{fig:nonren}
\end{center}
\end{figure}
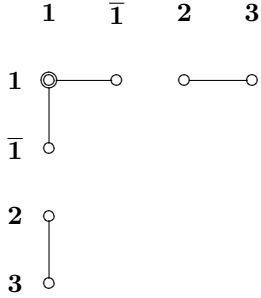

\begin{figure}[t]
\begin{center}
\begin{tabular}{c}
\begin{xy} 0;<3mm,0mm>:<0mm,3mm>::0;0,
,(3,0)*{{\bf 1}}
,(6,0)*{{\bf \bar 1}}
,(9,0)*{{\bf 2}}
,(12,0)*{{\bf 3}}
,(3,-3)*\cir(0.3,0){}
,(6,-3)*\cir(0.3,0){}
,(9,-3)*\cir(0.3,0){}
,(9.2,-3);(11.8,-3)**\dir{-}
,(3.2,-3);(5.8,-3)**\dir{-}
,(12,-3)*\cir(0.3,0){}
\end{xy}
\end{tabular}
\caption{The projected Krajewski diagram of Figure \ref{fig:nonren}.}
\label{fig:nonren-proj}
\end{center}
\end{figure}
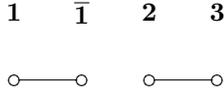

Note that, if renormalizable, the asymptotically expanded spectral action is not necessarily {\it multiplicatively} renormalizable, since the coefficients in front of the counterterms might be different for different factors in the gauge algebra, such as $\u(1)$, $\su(2)$ and $\su(3)$ for the Standard Model. 
This is in contrast with the classical action $\L_{M \times F}$ where there is a typical {\it unification} of couplings for all factors of the gauge algebra. This suggests that one takes the spectral action $\L_{M \times F}$ (plus gauge fixing) as a starting point for the renormalization group flow to then run the action to arbitrary energy scales. This is in concordance with the interpretation proposed in \cite{CC96}, and defines the starting point for a derivation of physical predictions. This approach led in \cite{CCM07,CM07} in the context of the Standard Model ---and assuming a big desert up to GUT-scale--- to a predicted Higgs mass around 168 GeV. The exclusion results at this value by Tevatron and the LHC suggests that one looks for other models within the noncommutative setup, describing physics at higher energies. 
A combination of the present renormalization conditions on almost-commutative geometries with experimental input is a modest first step towards finding such theories that go beyond the Standard Model. It might also indicate what more exotic (not almost-commutative) geometries will arise at higher energies, inevitably mixing the gauge theoretical content with the geometry of the background spacetime.

\newcommand{\noopsort}[1]{}\def\cprime{$'$}

\end{document}